\documentclass[aps,prl,preprint,tightenlines,superscriptaddress,showpacs,byrevtex,nobibnotes,refmerge]{revtex4}
\usepackage{graphicx} 
\usepackage{dcolumn}  

\def    \ra             {\rightarrow}
\def    \dd             {$D^0$-$\overline{D}{}^{\,0}$}
\def    \Agamma         {$A^{}_\Gamma$}
\def\simge{\mathrel{%
   \rlap{\raise 0.511ex \hbox{$>$}}{\lower 0.511ex \hbox{$\sim$}}}}
\def\simle{\mathrel{
   \rlap{\raise 0.511ex \hbox{$<$}}{\lower 0.511ex \hbox{$\sim$}}}}

\begin{document}



\preprint{\vbox{ \hbox{   }
                 \hbox{BELLE-CONF-347}
}}

\title{ \quad\\[0.5cm]  Measurement of the \dd\ lifetime 
difference using $D^0\rightarrow K\pi/KK$ decays}




\affiliation{Aomori University, Aomori}
\affiliation{Budker Institute of Nuclear Physics, Novosibirsk}
\affiliation{Chiba University, Chiba}
\affiliation{Chuo University, Tokyo}
\affiliation{University of Cincinnati, Cincinnati, Ohio 45221}
\affiliation{University of Frankfurt, Frankfurt}
\affiliation{Gyeongsang National University, Chinju}
\affiliation{University of Hawaii, Honolulu, Hawaii 96822}
\affiliation{High Energy Accelerator Research Organization (KEK), Tsukuba}
\affiliation{Hiroshima Institute of Technology, Hiroshima}
\affiliation{Institute of High Energy Physics, Chinese Academy of Sciences, Beijing}
\affiliation{Institute of High Energy Physics, Vienna}
\affiliation{Institute for Theoretical and Experimental Physics, Moscow}
\affiliation{J. Stefan Institute, Ljubljana}
\affiliation{Kanagawa University, Yokohama}
\affiliation{Korea University, Seoul}
\affiliation{Kyoto University, Kyoto}
\affiliation{Kyungpook National University, Taegu}
\affiliation{Institut de Physique des Hautes \'Energies, Universit\'e de Lausanne, Lausanne}
\affiliation{University of Ljubljana, Ljubljana}
\affiliation{University of Maribor, Maribor}
\affiliation{University of Melbourne, Victoria}
\affiliation{Nagoya University, Nagoya}
\affiliation{Nara Women's University, Nara}
\affiliation{National Kaohsiung Normal University, Kaohsiung}
\affiliation{National Lien-Ho Institute of Technology, Miao Li}
\affiliation{Department of Physics, National Taiwan University, Taipei}
\affiliation{H. Niewodniczanski Institute of Nuclear Physics, Krakow}
\affiliation{Nihon Dental College, Niigata}
\affiliation{Niigata University, Niigata}
\affiliation{Osaka City University, Osaka}
\affiliation{Osaka University, Osaka}
\affiliation{Panjab University, Chandigarh}
\affiliation{Peking University, Beijing}
\affiliation{Princeton University, Princeton, New Jersey 08545}
\affiliation{RIKEN BNL Research Center, Upton, New York 11973}
\affiliation{Saga University, Saga}
\affiliation{University of Science and Technology of China, Hefei}
\affiliation{Seoul National University, Seoul}
\affiliation{Sungkyunkwan University, Suwon}
\affiliation{University of Sydney, Sydney NSW}
\affiliation{Tata Institute of Fundamental Research, Bombay}
\affiliation{Toho University, Funabashi}
\affiliation{Tohoku Gakuin University, Tagajo}
\affiliation{Tohoku University, Sendai}
\affiliation{Department of Physics, University of Tokyo, Tokyo}
\affiliation{Tokyo Institute of Technology, Tokyo}
\affiliation{Tokyo Metropolitan University, Tokyo}
\affiliation{Tokyo University of Agriculture and Technology, Tokyo}
\affiliation{Toyama National College of Maritime Technology, Toyama}
\affiliation{University of Tsukuba, Tsukuba}
\affiliation{Utkal University, Bhubaneswer}
\affiliation{Virginia Polytechnic Institute and State University, Blacksburg, Virginia 24061}
\affiliation{Yokkaichi University, Yokkaichi}
\affiliation{Yonsei University, Seoul}
  \author{K.~Abe}\affiliation{High Energy Accelerator Research Organization (KEK), Tsukuba} 
  \author{K.~Abe}\affiliation{Tohoku Gakuin University, Tagajo} 
  \author{N.~Abe}\affiliation{Tokyo Institute of Technology, Tokyo} 
  \author{R.~Abe}\affiliation{Niigata University, Niigata} 
  \author{T.~Abe}\affiliation{High Energy Accelerator Research Organization (KEK), Tsukuba} 
  \author{I.~Adachi}\affiliation{High Energy Accelerator Research Organization (KEK), Tsukuba} 
  \author{Byoung~Sup~Ahn}\affiliation{Korea University, Seoul} 
  \author{H.~Aihara}\affiliation{Department of Physics, University of Tokyo, Tokyo} 
  \author{M.~Akatsu}\affiliation{Nagoya University, Nagoya} 
  \author{M.~Asai}\affiliation{Hiroshima Institute of Technology, Hiroshima} 
% ...
\collaboration{The Belle Collaboration}


\date{\today}

\begin{abstract}
We report a preliminary measurement of the \dd\ mixing parameter
$y_{\rm CP}$ and the CP-violating parameter \Agamma\ using
the decay $D^{*+}\ra D^0\pi^+$ followed by $D^0 \ra K^- \pi^+$ and 
$D^0 \ra K^+ K^-$. The results are obtained from a $158\,{\rm
fb}^{-1}$ data sample collected near the $\Upsilon(4S)$ resonance
with the Belle detector at the KEKB asymmetric energy $e^+ e^-$ collider.
\end{abstract}


\maketitle


{\renewcommand{\thefootnote}{\fnsymbol{footnote}}}
\setcounter{footnote}{0}

\section{Introduction}

The rate of \dd\ mixing, which has not yet been observed, would provide 
an important window on new physics. The contribution of box diagrams 
is very small due to GIM suppression, and the mixing rate is believed
to be dominated by long-distance processes. These processes are 
themselves suppressed by $SU(3)$-flavor symmetry.
The mixing rate is usually expressed in terms of the parameters
$x\equiv\Delta m/\Gamma$ and $y\equiv\Delta\Gamma/(2\Gamma)$,
where $\Delta m$ and $\Delta \Gamma$ are the mass- and width-differences 
of the physical states, and $\Gamma$ is the average decay width. 
Numerous predictions for $x$ and $y$ exist~\cite{nelson}; a recent 
calculation gives $x\!\sim\!y\!\sim\!10^{-3}$~\cite{bigi-uraltsev}.
Non-Standard-Model (SM)  processes are expected to enhance $x$ relative 
to $y$, and measuring $x\gg y$ would be a strong indication of 
new physics~\cite{bigi-uraltsev,falk}.

In this paper we report a measurement of the parameter 
$$y^{}_{\rm CP}\ \equiv \frac{\tau(K^- \pi^+)}{\tau(K^+ K^-)} - 1\,,$$
which is equal to $y$ in the limit where CP is conserved.
If CP is violated, the widths of $D^0$ and $\overline{D}{}^{\,0}$ decays 
into the CP eigenstate $K^+ K^-$ differ.
These widths can be expressed as~\cite{petrov}
\begin{eqnarray*}
\hat{\Gamma}(D^0 \ra K^+ K^-) & = &  
\Gamma \left[\,1 + R_m (y \cos{\phi} - x \sin{\phi})\right] \\
\hat{\Gamma}(\overline{D}{}^{\,0} \ra K^+ K^-) & = &  
\Gamma \left[\,1 + R_m^{-1} (y \cos{\phi} + x \sin{\phi})\right]\,,
\end{eqnarray*}
where $R^{}_m$ and $\phi$ are CP-violating parameters defined in~\cite{petrov}.
CP violation in mixing results in $R_m \neq 1$, and CP violation
due to interference between mixed and unmixed decay amplitudes 
results in $\phi \neq 0$. We investigate whether CP violation 
is present by measuring the quantity
$$A^{}_\Gamma\ \equiv\ 
\frac{\hat{\Gamma}(D^0 \ra K K) - \hat{\Gamma}(\overline{D}{}^{\,0} \ra K K)}
{\hat{\Gamma}(D^0 \ra K K) + \hat{\Gamma}(\overline{D}{}^{\,0} \ra K K)}\ \approx\  
\left(\frac{R^{}_m-R^{-1}_m}{R^{}_m+R^{-1}_m}\right)
y\cos{\phi}-x\sin{\phi}\,.$$
The SM predicts $A^{}_\Gamma\simle 10^{-4}$~\cite{bigi-sanda}; 
a large value would indicate physics beyond the~SM.

Our measurement of $y_{CP}$ is obtained by measuring 
the lifetime difference between $D^0 \ra K^- \pi^+$ and $D^0 \ra K^+ K^-$ decays.
(Throughout this paper, charge-conjugate modes are included unless 
noted otherwise.) 
Our measurement of \Agamma\ is obtained by measuring the lifetime difference 
between $D^0 \ra K^+ K^-$ and $\overline{D}{}^{\,0} \ra K^+ K^-$ decays.
The analysis is based on 158~fb$^{-1}$ of data, and
all results are preliminary.
To identify the flavor of the $D^0$ and also reduce background, 
we require that all $D^0$'s originate from 
$D^{*+}\ra D^0\pi^+$ decays.

\section{Event selection}

The Belle detector is a large-solid-angle magnetic
spectrometer that
consists of a three-layer silicon vertex detector (SVD),
a 50-layer central drift chamber (CDC), an array of
aerogel threshold \v{C}erenkov counters (ACC), 
a barrel-like arrangement of time-of-flight
scintillation counters (TOF), and an electromagnetic calorimeter
comprised of CsI(Tl) crystals (ECL) located inside 
a super-conducting solenoid coil that provides a 1.5~T
magnetic field.  An iron flux-return located outside of
the coil is instrumented to detect $K_L^0$ mesons and to identify
muons (KLM).  The detector
is described in detail elsewhere~\cite{Belle}.

We select well-measured charged tracks that have impact parameters 
less than 1.0~cm in the radial direction and
less than 2.0~cm along the beam direction ($z$) with respect
to the interaction point (IP). The $D^0$ daughter tracks are 
required to have at least two hits in each of the $r$-$\phi$ 
and $z$-measuring SVD layers. This requirement is not applied 
to the slow pion from $D^{*+} \ra D^0 \pi^+$ decays.

For charged particle identification, information from the CDC, 
TOF, and ACC subsystems is combined to form an overall hadron likelihood:
$$ \mathcal{L}(h)\ =\ \mathcal{L}^{\rm ACC}(h) \times 
\mathcal{L}^{\rm CDC}(h) \times \mathcal{L}^{\rm TOF}(h)~,$$ 
where $h$ stands for $\pi$ or $K$. Charged particles are
identified as pions or kaons using the likelihood ratios:
$$\mathcal{P}(K)\ =\ \frac{\mathcal{L}(K)}{\mathcal{L}(K)+\mathcal{L}(\pi)}\,,
\hskip0.50in
\mathcal{P}(\pi)\ =\ \frac{\mathcal{L}(\pi)}{\mathcal{L}(K)+\mathcal{L}(\pi)} 
\ =\ 1 - \mathcal{P}(K)~.$$
Kaon candidates are required to have $\mathcal{P}(K)>0.6$.
This requirement is 88\% efficient and has a pion 
misidentification probability of 8\%. Pion candidates 
are required to satisfy $\mathcal{P}(K) < 0.9$.

Both continuum and $\Upsilon(4S)$ data are used in this analysis.
Each event is required to contain at least one $D^{*+}$ meson 
decaying to $D^0 \pi^+$. To remove $D^{*+}$'s resulting from 
$B$ meson decays, the $D^{*+}$ is required to have a momentum in the
$e^+ e^-$ center-of-mass frame (CMS) greater than 2.5~$\mathrm{GeV}/c$.
$D^0$ meson candidates are reconstructed from $D^0 \ra K^- \pi^+$ and $D^0 \ra K^+ K^-$ decays.
All $D^0$ candidates must satisfy $|M(D^0)_{\rm fit} - M(K^- \pi^+/K^+ K^-)| < 2 \sigma_{\rm fit}$, 
where $M(D^0)_{\rm fit}$
and $\sigma_{\rm fit}$ are the mean and standard deviation 
of a Gaussian fit to the $D^0$ mass peak in data.
These parameters are calculated 
independently for each $D^0$ decay mode in order to 
help keep the $K^- \pi^+$ and $K^+ K^-$
vertex resolution functions unbiased by the $D^0$ mass requirement.
From the fits, $\sigma_{\rm fit}$ is found to be 5.1~$\mathrm{MeV}/c^2$ 
for $D^0 \ra K^- \pi^+$ and 4.5~$\mathrm{MeV}/c^2$ for $D^0 \ra K^+K^-$ 
(see Table~\ref{tab-datamass}). 
The momentum of the slow $\pi^+$ from $D^{*+} \ra D^0 \pi^+$ decays 
is obtained by refitting using the beam position constraint,
i.e., the track must project back to the IP region.
The invariant mass of the $D^{*+}$ candidates
after slow $\pi^+$ refitting is required to satisfy
$|M(D^0 \pi^+) - M(K^- \pi^+/K^+ K^-) + 
M_{D^0}^{\rm PDG} - M_{D^*}^{\rm PDG}| < 1.0~\mathrm{MeV}/c^2$.

\section{$\mathbf{D^0}$ vertex fit and decay time reconstruction}

The proper time between the production and decay of the $D^0$ candidates 
is determined by projecting the momentum $\vec{P}$ and the flight 
length $\vec{\ell}$ onto the $xy$ plane transverse to the beam axis:
$$t_{D^0}\ =\ \frac{\vec{\ell}_{xy} \cdot \vec{P}_{xy}}{P_{xy}^2} 
\ \times\ \left(\frac{m_{D^0}}{c}\right)\,.$$
The momentum $\vec{P}$ is the vector sum of the $K$ and $\pi$ momenta,
and $\vec{\ell}$ is the displacement vector between the 
production and decay vertices.
Information from the $z$ projections is not included
due to the large uncertainty in the $z$ component of the IP.

\begin{figure}[hbt]
\includegraphics[width=0.48\textwidth]{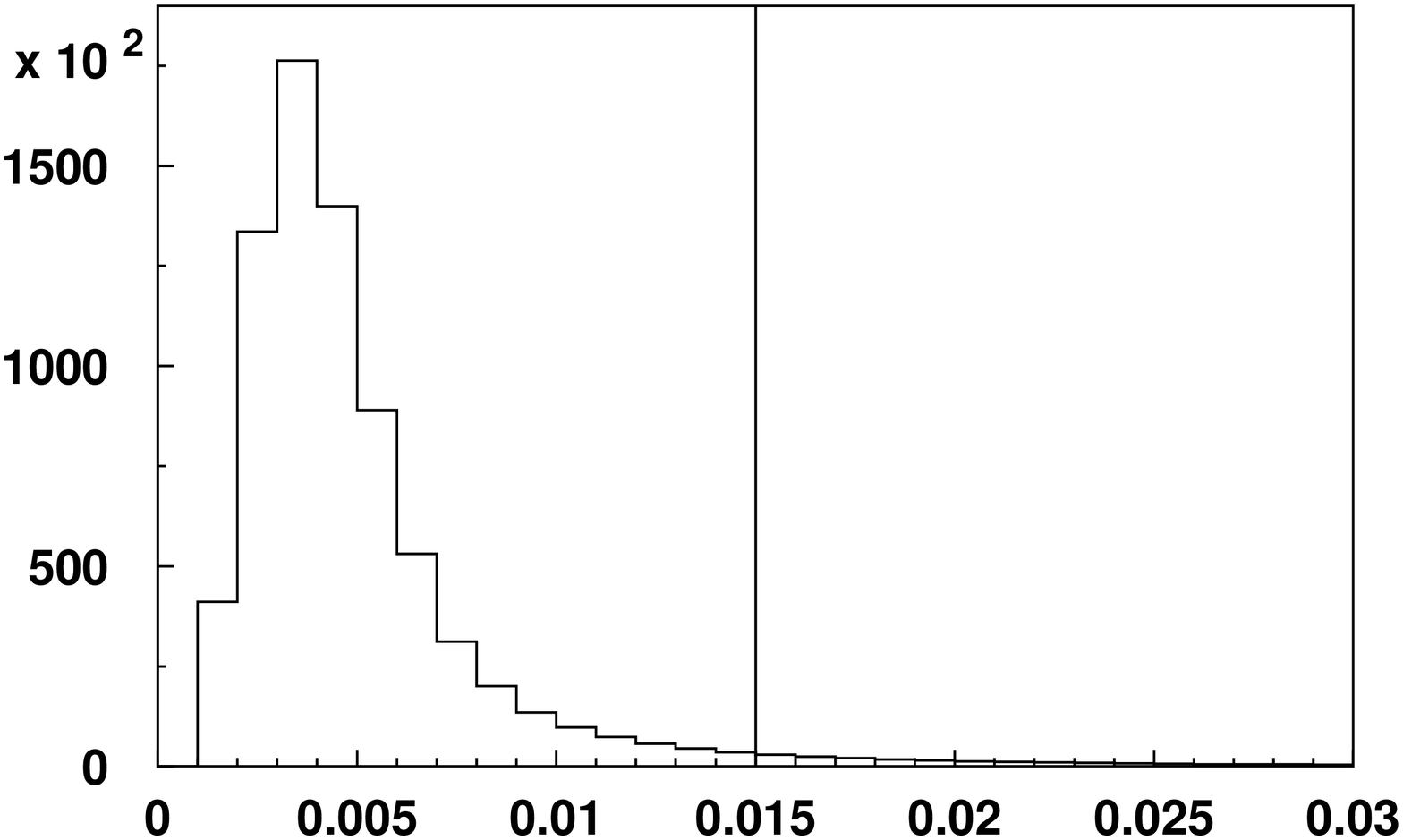}
\includegraphics[width=0.48\textwidth]{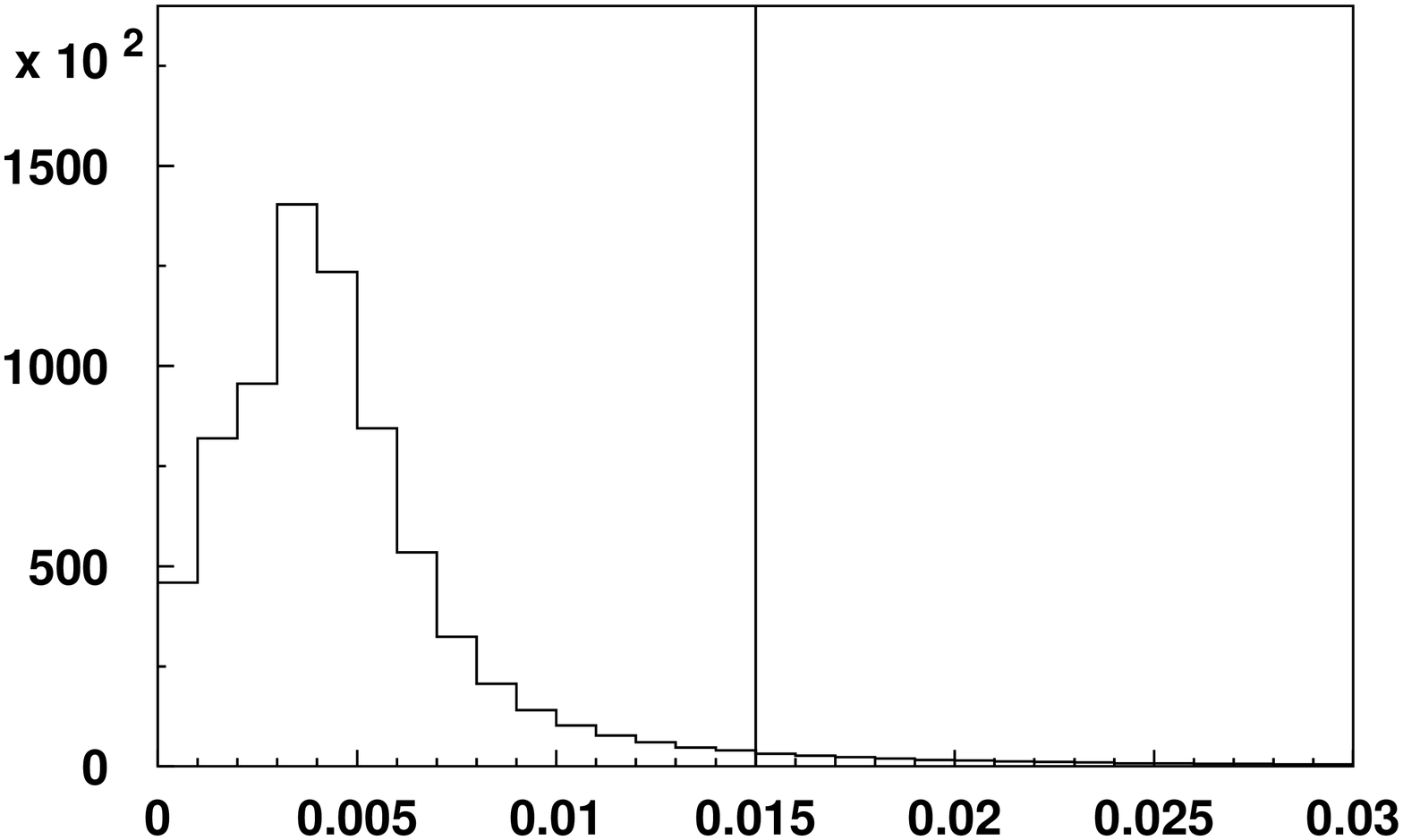}
\caption{Distributions of decay vertex errors $\sigma_x$ (left plot) 
and $\sigma_y$ (right plot) returned by the $D^0$ vertex fit after 
the IP position correction. The horizontal scale is in~cm. Vertical lines
indicate the position of the fit quality cut: $\sigma_{x, y} < 150~\mu$m.}
\label{sigxy}
\end{figure}

To improve the spatial resolution, the vertex fit of
$D^0 \ra K^- \pi^+ (K^+ K^-)$ decays is done in three steps. 
First, the decay point is fit using only information from the 
$D^0$ daughter tracks. Then, the $D^0$ production vertex is 
found by fitting the reconstructed $D^0$ 
momentum vector with the IP region, whose profile is 
obtained from a separate study of the full data set.
Finally, the $D^0$ decay vertex 
is corrected using information from the $D^0$ production vertex fit.
No mass constraint is applied to the $D^0$ candidates during the vertex fit
to avoid mode-dependent bias that could result from a correlation between
the invariant mass and a global momentum correction procedure.
The error in the position of the $D^0$ decay vertex
obtained from the fit is required to be less than 150~$\mu$m 
in each of the $x$ and $y$ directions. Fig.~\ref{sigxy} 
illustrates the high efficiency of this cut. This procedure 
produces typical $D^0$ decay vertex resolutions of 
$45~\mu$m in both $x$ and $y$. 

\begin{figure}[htb]
\includegraphics[width=0.48\textwidth]{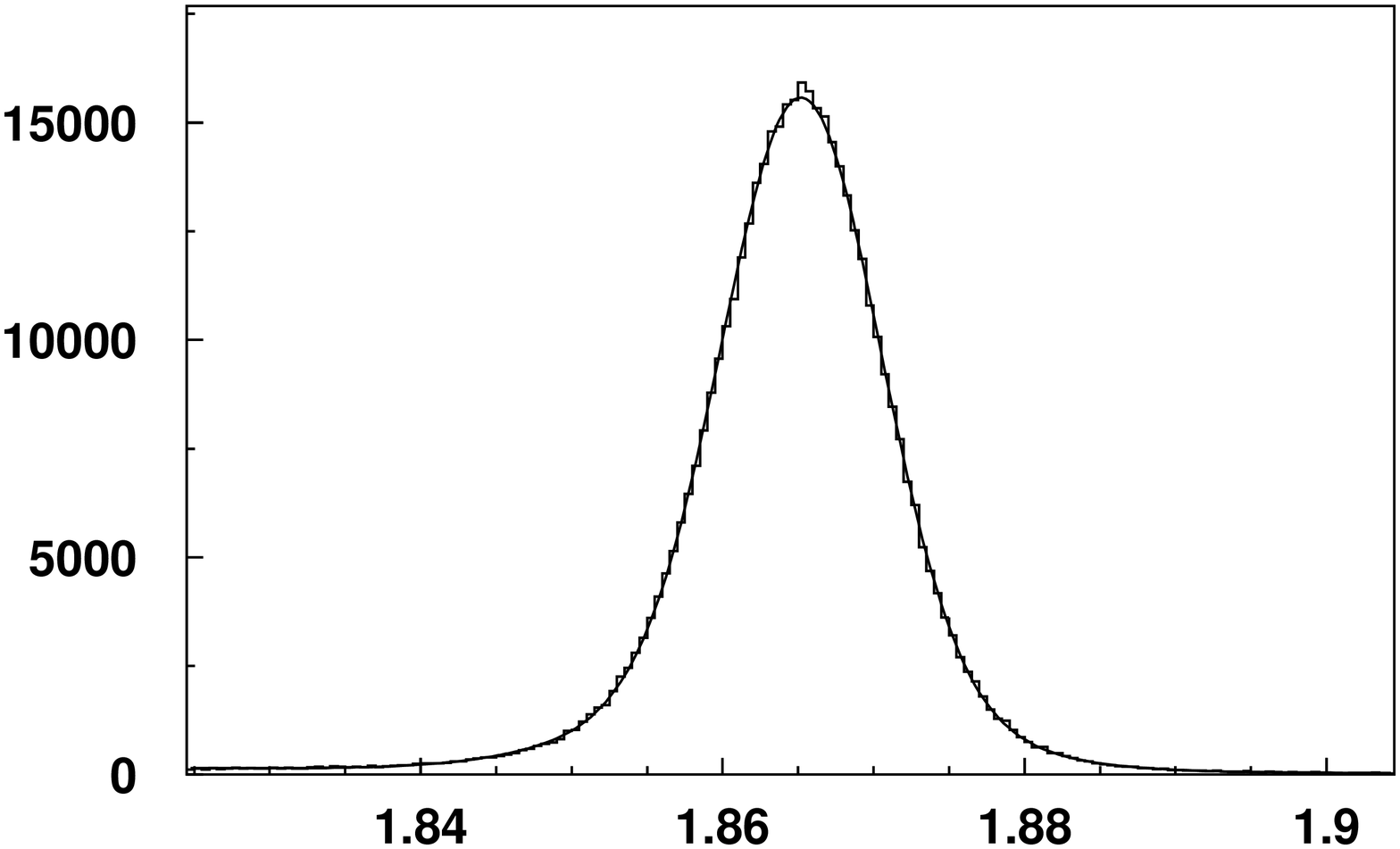}
\includegraphics[width=0.48\textwidth]{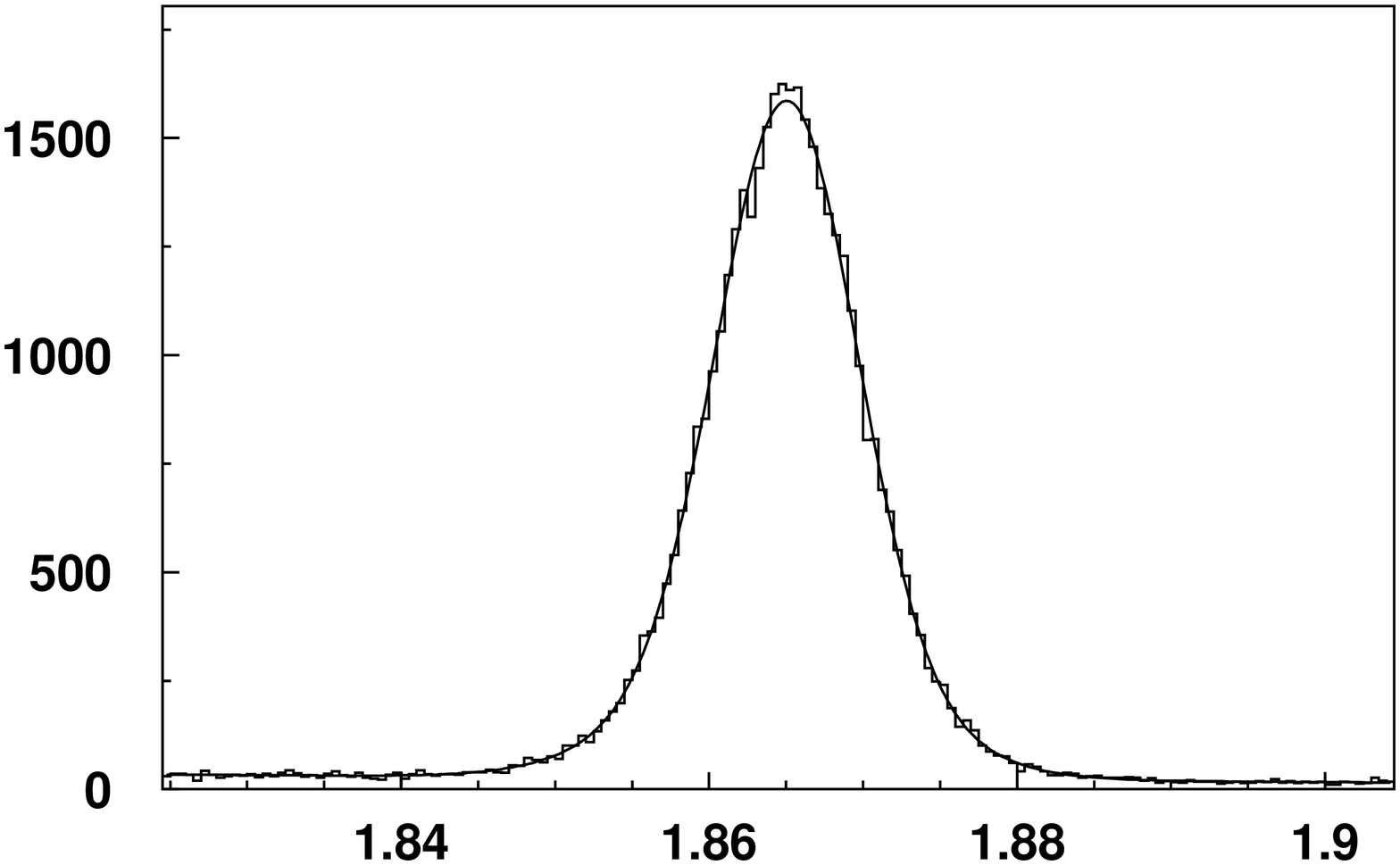}
\caption{Invariant mass spectra for $D^0 \ra K^- \pi^+$ candidates
(left plot) and $D^0 \ra K^+ K^-$ candidates (right plot). The horizontal
scale is in~GeV$/c^2$. The bin size is 0.5~MeV$/c^2$ and the  
solid curves show the fit results.}
\label{fig-datamass}
\end{figure}
The invariant mass spectra of the selected $D^0 \ra K^- \pi^+$ and 
$D^0 \ra K^+ K^-$ candidates are shown in Fig.~\ref{fig-datamass}. 
The distributions are fit with a sum of two Gaussians representing 
the signal and a first-order polynomial representing the background. 
The results of the fit are summarized in Table~\ref{tab-datamass}.

\begin{table}[htb]
\caption{Invariant mass spectra fit results.}
\label{tab-datamass}
\begin{tabular}
{@{\hspace{0.3cm}}c@{\hspace{0.3cm}} @{\hspace{0.3cm}}c@{\hspace{0.3cm}}
@{\hspace{0.3cm}}c@{\hspace{0.3cm}} @{\hspace{0.3cm}}c@{\hspace{0.3cm}}
c@{\hspace{0.3cm}}}
\hline \hline
Mode & \# events & main Gaussian $\sigma$ & main Gaussian fraction 
& Signal purity \\
     &      &  (MeV/$c^2$) & (\%) & (\%) \\
\hline
$K^- \pi^+$ & 448000 & $5.1$ & 78.4 & 99.1 \\
\hline
$K^+ K^-$ & 36480 & $4.5$ & 74.0 & 97.6 \\
\hline \hline
\end{tabular}
\end{table}

\section{Resolution function study}

\begin{figure}[htb]
\includegraphics[width=0.48\textwidth]{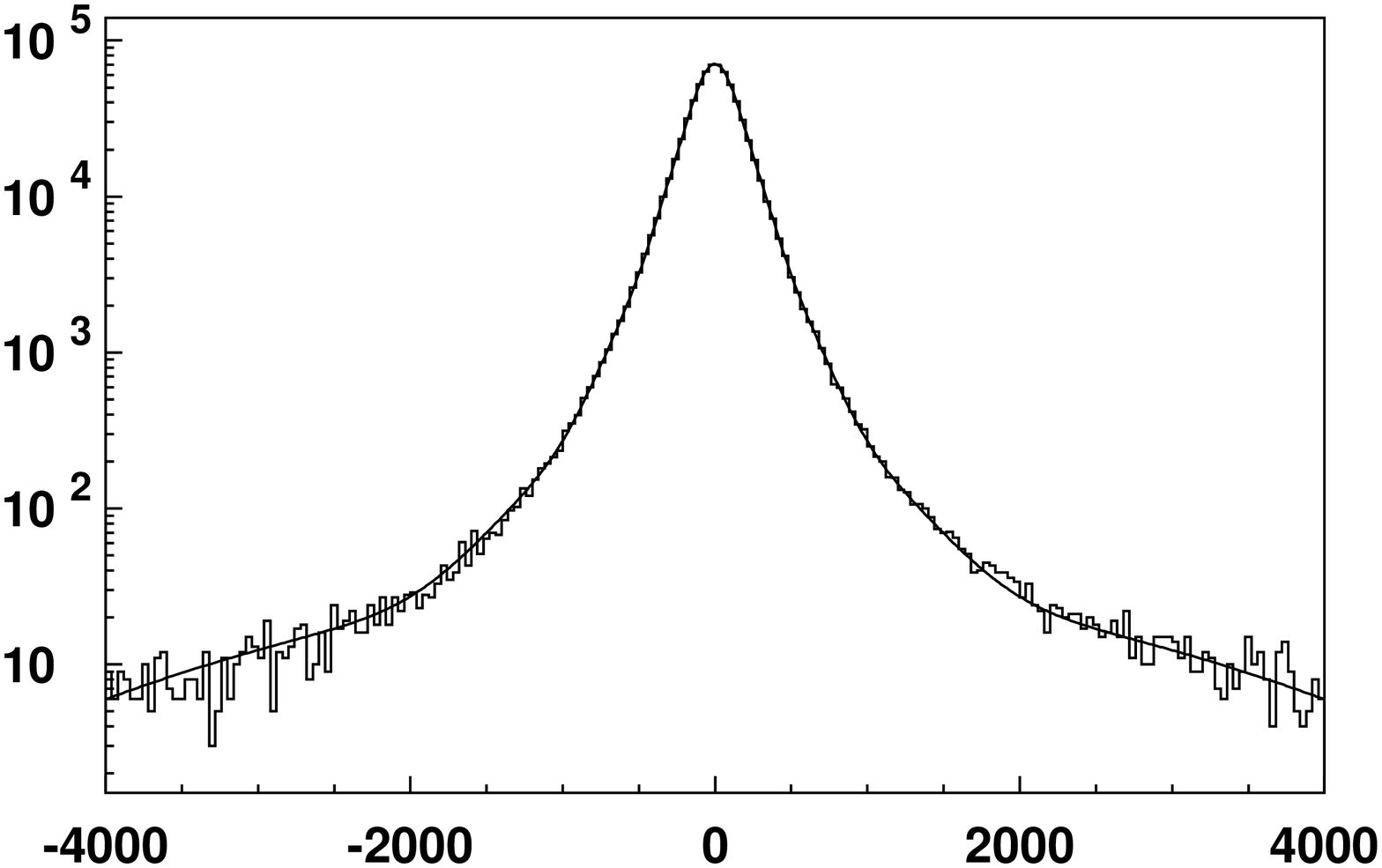}
\includegraphics[width=0.48\textwidth]{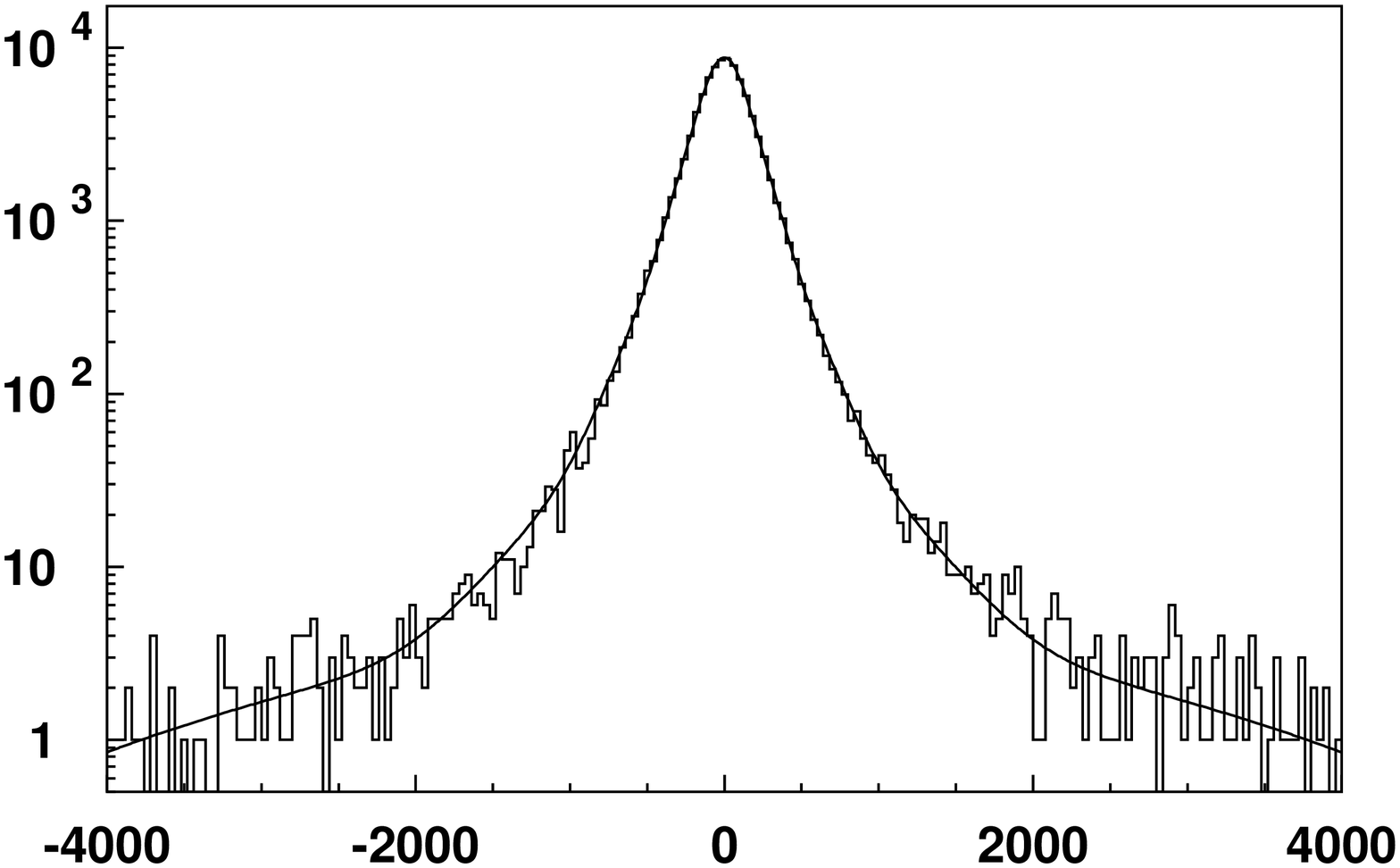}
\caption{Proper time resolution functions for $D^0 \ra K^- \pi^+$ 
(left plot) and $D^0 \ra K^+ K^-$ (right plot) obtained from MC 
simulation. The horizontal scale is in~fs. The bin size of the 
data points is 40~fs and the solid curves show the results of 
the fit. 
(The fit is performed using a bin size of 10~fs.)
}
\label{fig-res}
\end{figure}

The proper time distribution is represented by an exponential 
function with time constant $\tau^{}_D$ 
convolved with the detector vertex resolution. The properties 
of the detector resolution function are studied using Monte Carlo (MC) 
simulation of the signal decay modes. Approximately 
$7\cdot 10^5$ $D^0 \ra K^- \pi^+$ decays and $1\cdot 10^5$ 
$D^0 \ra K^+ K^-$ decays were simulated, corresponding to 
two times the data sample and three times the data sample, respectively.

Fig.~\ref{fig-res} shows the difference between the generated and
reconstructed decay times of the $D^0$ mesons. The distributions are fit 
with a sum of five Gaussians constrained to a common central value. 
The results of the fit are summarized in Table~\ref{tab-res};
the confidence level (CL) is 22.2\%.
Due to similar kinematics, the $K^+ K^-$ decay mode is fit with
the same resolution function as that used for $K^- \pi^+$.
However, to account for small differences between $K^+ K^-$ 
and $K^- \pi^+$, the width of each Gaussian component in 
the $K^+ K^-$ fit is multiplied by a common scale factor
$\alpha$. Fitting first the $K^- \pi^+$ distribution and 
then the $K^+ K^-$ distribution, we obtain $\alpha = 1.043 \pm 0.004$.
The CL for the $K^+ K^-$ fit is 71.3\%. The common central values of 
the Gaussians are found to be $X_0 = -1.51 \pm 0.22$~fs for $K^- \pi^+$
and $X_0 = -1.57 \pm 0.63$~fs for $K^+ K^-$.

In the lifetime difference fits for both MC and data, 
the relative fractions of the Gaussian components of
the resolution function are fixed to the values 
determined in the above study.
The central value of the $K^+ K^-$ resolution function 
is fixed to that of the $K^- \pi^+$ resolution function,
which is consistent with the results above. The possible bias due 
to this assumption is included in the systematic error.


\begin{table}[htb]
\caption{ $D^0 \ra K^- \pi^+$ proper time resolution function fit results.}
\label{tab-res}
\begin{tabular}
{@{\hspace{1.0cm}}c@{\hspace{1.0cm}} @{\hspace{1.0cm}}c@{\hspace{1.0cm}}
@{\hspace{1.0cm}}c@{\hspace{1.0cm}}}
\hline \hline
Fit parameter & Fraction (\%) & Value (fs) \\
\hline
$\sigma_1$ & 26.1 & $95.1 \pm 1.3$ \\
\hline
$\sigma_2$ & 50.4 & $177.0 \pm 2.2$ \\
\hline
$\sigma_3$ & 19.8 & $328.7 \pm 7.4$ \\
\hline
$\sigma_4$ & 3.1 & $675.7 \pm 24.9$ \\
\hline
$\sigma_5$ & 0.6 & $2199 \pm 95$ \\
\hline \hline
\end{tabular}
\end{table}

\section{MC study of lifetime difference fit}

\begin{figure}[htb]
\includegraphics[width=0.48\textwidth]{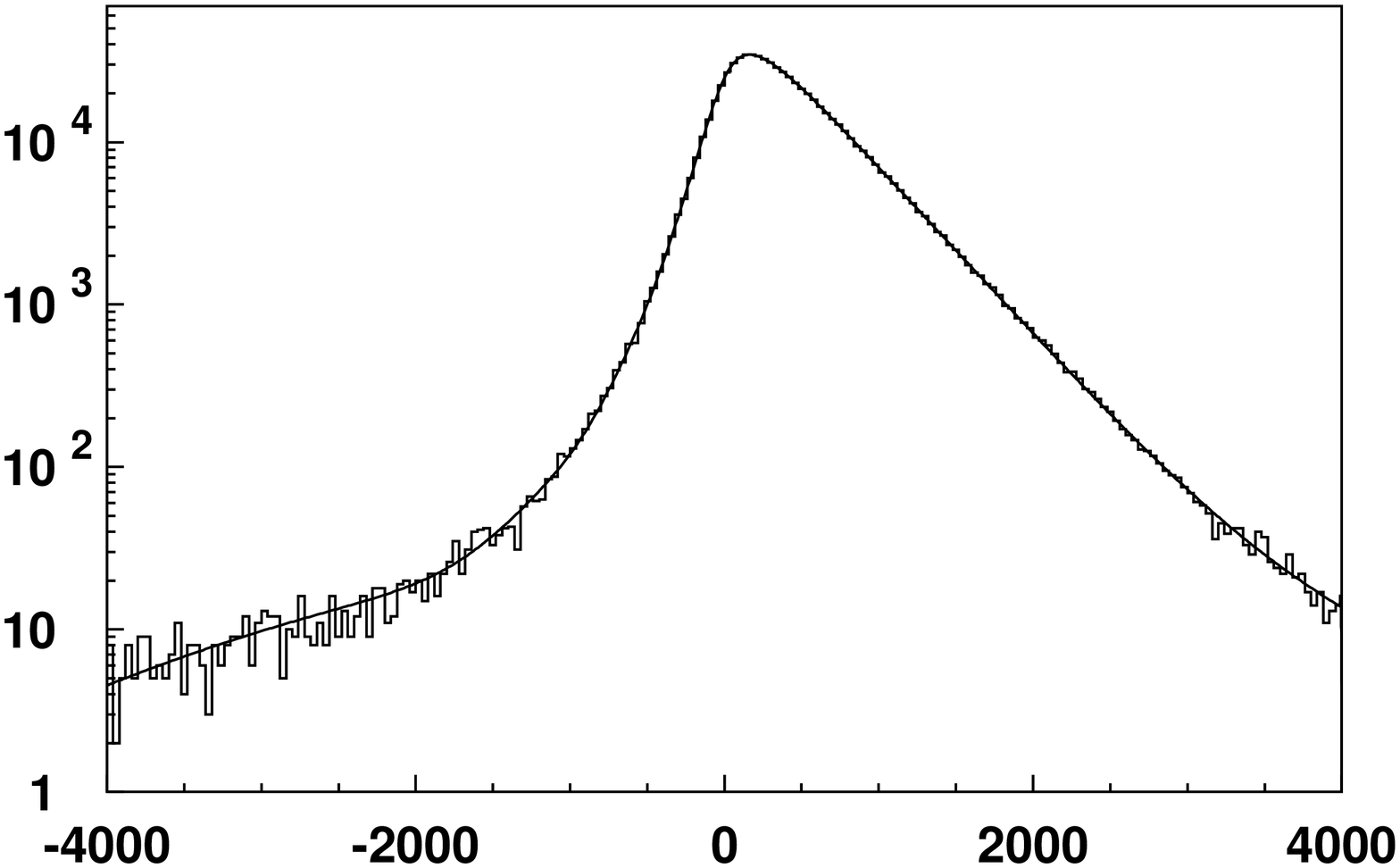}
\includegraphics[width=0.48\textwidth]{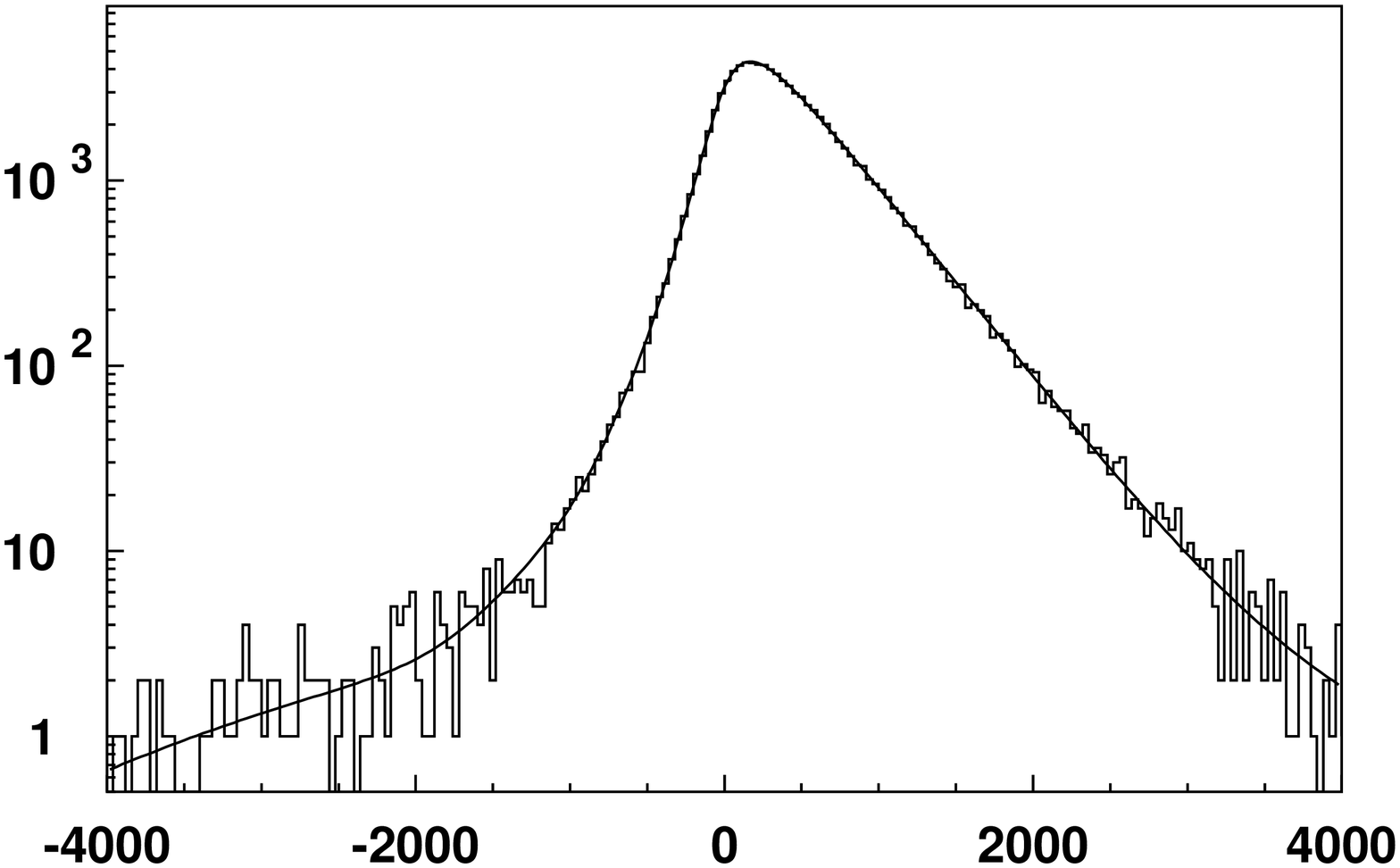}
\caption{Simultaneous lifetime difference fit for $D^0 \ra K^- \pi^+$ 
(left plot) and $D^0 \ra K^+ K^-$ (right plot) performed on MC-generated 
events. The horizontal scale is in~fs. The bin size of the data points 
is 40~fs and the solid curves show the results of the fit.
(The fit was done using a bin size of 10~fs.)
}
\label{fig-mcfit}
\end{figure}

The $D^0$ lifetime difference is obtained from
a simultaneous, binned, maximum likelihood fit 
to the measured proper lifetime distributions of
$D^0 \ra K^+ K^-$ and $D^0 \ra K^- \pi^+$ decays.
The consistency and quality of the  
fitting procedure is tested with MC events.
The lifetime distributions of $D^0\ra K\pi/KK$ decays were 
fit with an exponential function convolved with a resolution 
function of the form described above. The CL of the fit is 61\%. 
%
%
Fig.~\ref{fig-mcfit} and Table~\ref{tab-mcfit} show the results 
of the fit and a comparison to the parameters obtained
from the resolution function study. The parameters are 
seen to agree to within the quoted errors. The 
resolution function scale factor $\alpha$ determined from 
the fit is $1.042 \pm 0.007$, also in agreement with the 
result from the resolution function study. The $D^0$ 
lifetime obtained is $411.9 \pm 0.8$~fs, which is consistent 
with the generated value of 412.6~fs. The lifetime difference 
obtained is $0.0 \pm 1.7$~fs, which is consistent with the 
generated value of exactly 0~fs.

\begin{table}[htb]
\caption{Results of the simultaneous lifetime fit (MC) vs. standalone resolution function fit.}
\label{tab-mcfit}
\begin{tabular}
{@{\hspace{1.0cm}}c@{\hspace{1.0cm}} @{\hspace{1.0cm}}c@{\hspace{1.0cm}}
@{\hspace{1.0cm}}c@{\hspace{1.0cm}}} 
\hline \hline
Fit parameter & Resolution Function (fs) & Lifetime Fit (fs) \\
\hline
$\sigma_1$ & $95.1 \pm 1.3$ & $94.4 \pm 1.7$ \\
\hline
$\sigma_2$ & $177.0 \pm 2.2$ & $179.0 \pm 1.2$ \\
\hline
$\sigma_3$ & $328.7 \pm 7.4$ & $328.2 \pm 2.2$ \\
\hline
$\sigma_4$ & $675.7 \pm 24.9$ & $664.4 \pm 8.5$ \\
\hline
$\sigma_5$ & $2199 \pm 95$ & $2225 \pm 70$ \\
\hline
$X_0$ & $-1.51 \pm 0.22$ & $-0.95 \pm 0.54$ \\ 
\hline
\hline
$\alpha$ & $1.043 \pm 0.004$ & $1.042 \pm 0.007$ \\ 
\hline \hline
\end{tabular}
\end{table}

\section{Lifetime difference fit in DATA}

To fit the lifetime difference in data, one must include small
contributions from background processes. To include these,
the proper time distribution of background is obtained from events 
in the $D^0$ mass sidebands. The sidebands chosen begin 
$\pm25$~MeV$/c^2$ from $m^{}_{D^0}$ and cover the same total range
as the signal region ($4\sigma_{\rm fit}$).

\begin{figure}[htb]
\includegraphics[width=0.48\textwidth]{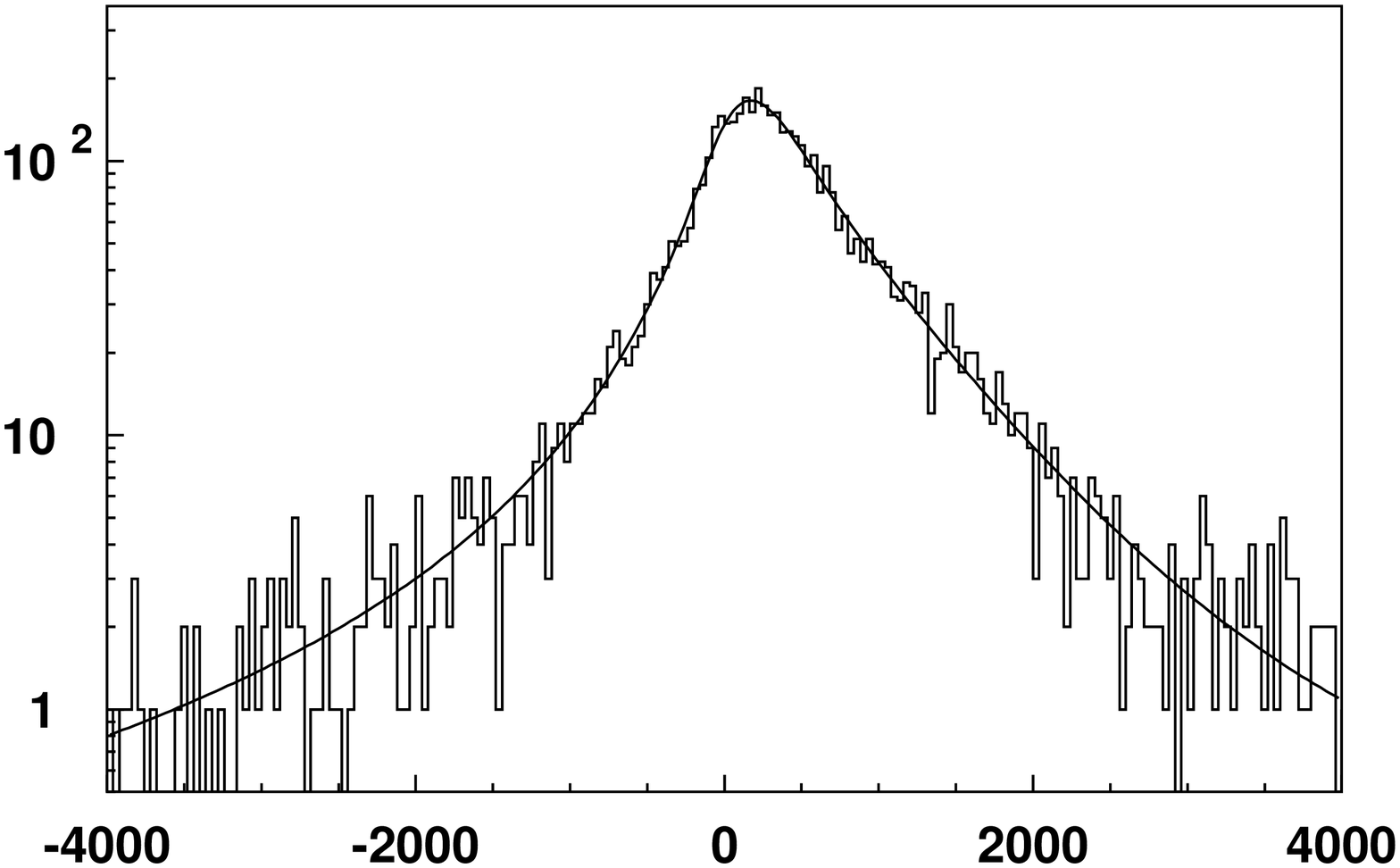}
\includegraphics[width=0.48\textwidth]{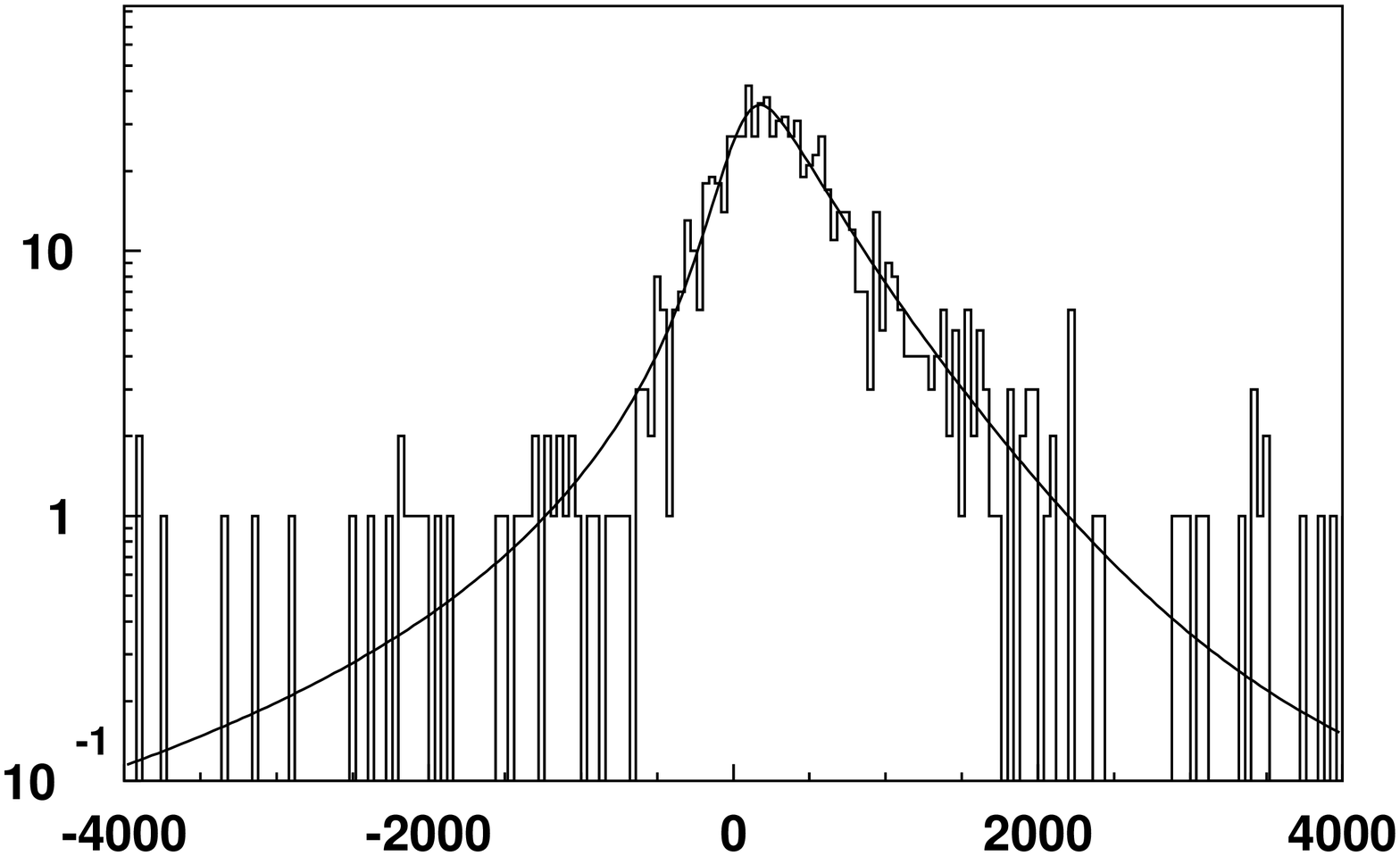}
\caption{Background proper time distributions for $D^0 \ra K^- \pi^+$ 
(left plot) and $D^0 \ra K^+ K^-$ (right plot). The horizontal
scale is in~fs. The bin size is 40~fs and the solid curves show 
the results of the fit.}
\label{fig-bkg}
\end{figure}

The background proper time distributions are parameterized as the sum of
an exponential convolved with a Gaussian and a Breit-Wigner function. 
The resulting spectra are shown in Fig.~\ref{fig-bkg}. The systematic 
uncertainties due to the background fitting procedure are evaluated 
by varying the fitting parameters. 
Since the background contribution
is small compared to the signal, the effect of the background 
magnitude and shape uncertainty is small (approximately $0.1\%$ in
$y_{CP}$). It was checked with MC simulation that the small fraction
of signal events remaining in the mass sidebands does not affect the $y_{CP}$ 
measurement above the quoted systematic error on background. The decay 
time function obtained for background is included (with normalization 
fixed) in the lifetime fit to the signal region.

\begin{figure}[htb]
\includegraphics[width=0.48\textwidth]{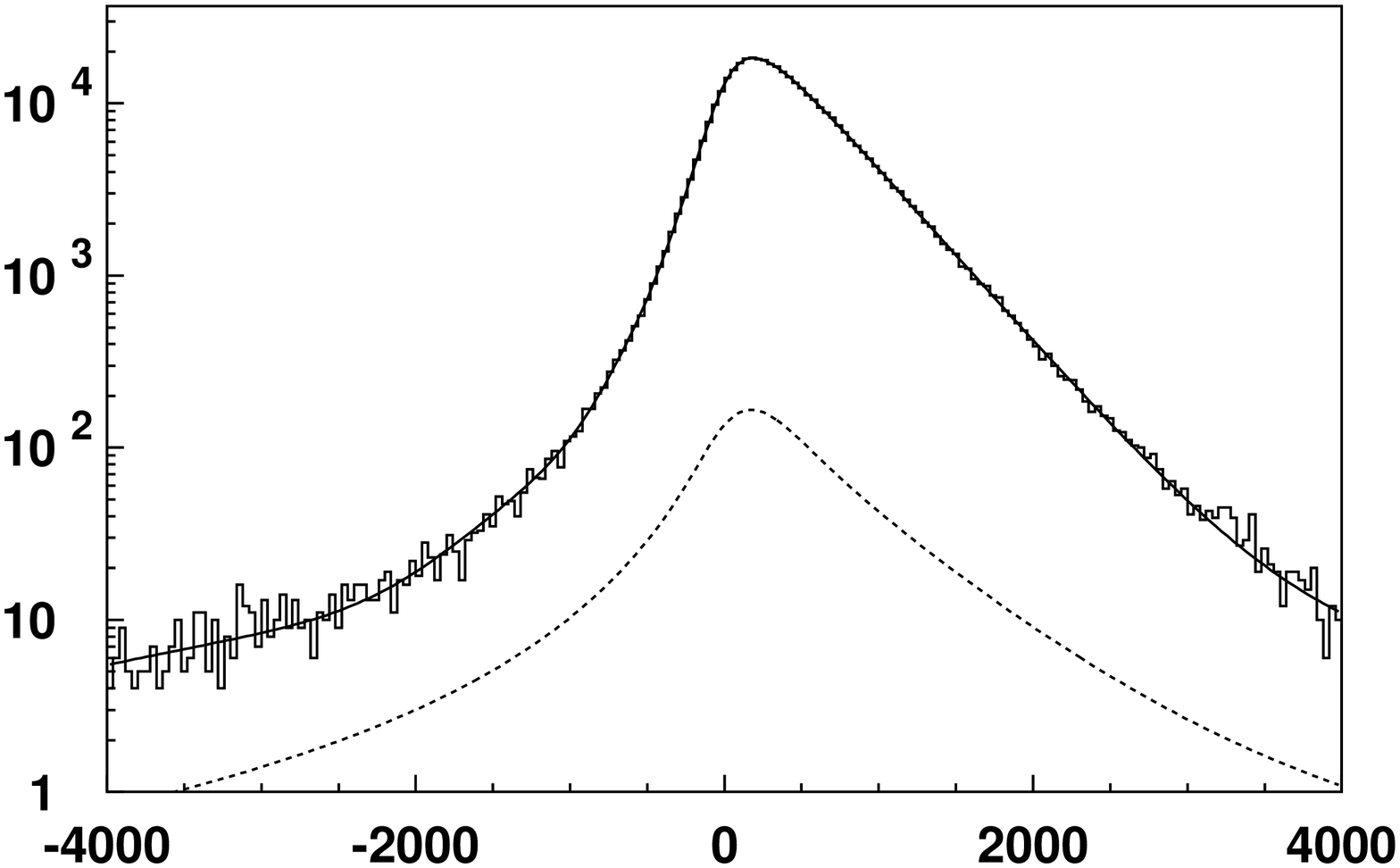}
\includegraphics[width=0.48\textwidth]{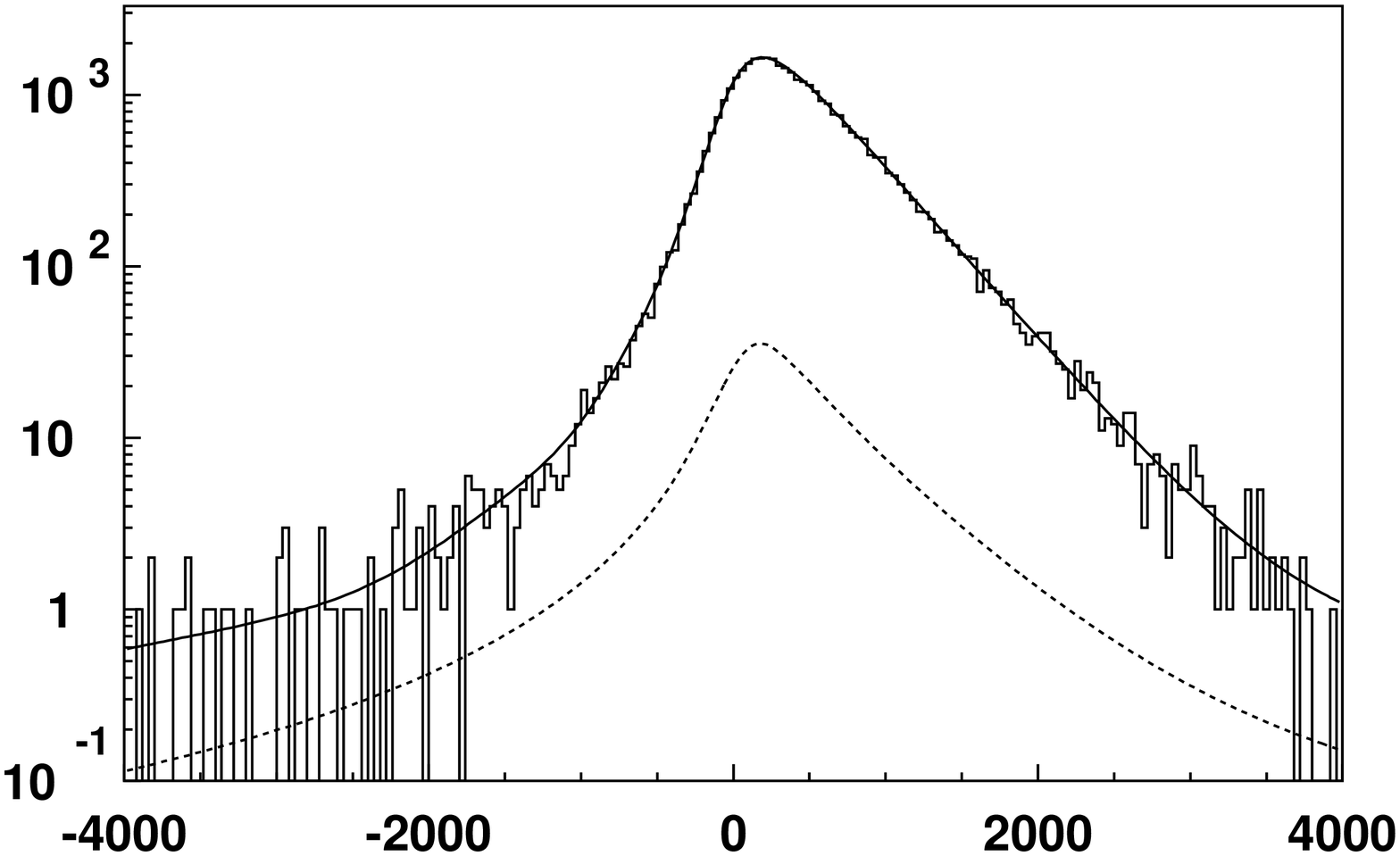}
\caption{Simultaneous lifetime difference fit to the data for 
$D^0 \ra K^- \pi^+$ (left plot) and $D^0 \ra K^+ K^-$ (right plot).
The horizontal scale is in~fs. The bin size is 40~fs and the solid 
curves show the results of the fit. The dashed curves show the 
background component included in the fit.}
\label{fig-datafit}
\end{figure}

\begin{table}[htb]
\caption{Results of the simultaneous lifetime fit (data) vs. standalone resolution function fit.}
\label{tab-datafit}
\begin{tabular}
{@{\hspace{1.0cm}}c@{\hspace{1.0cm}} @{\hspace{1.0cm}}c@{\hspace{1.0cm}}
@{\hspace{1.0cm}}c@{\hspace{1.0cm}}} 
\hline \hline
Fit parameter & Resolution Function (fs) & Lifetime Fit (fs) \\
\hline
$\sigma_1$ & $95.1 \pm 1.3$ & $112.3 \pm 2.6$ \\
\hline
$\sigma_2$ & $177.0 \pm 2.2$ & $198.2 \pm 1.8$ \\
\hline
$\sigma_3$ & $328.7 \pm 7.4$ & $378.0 \pm 3.2$ \\
\hline
$\sigma_4$ & $675.7 \pm 24.9$ & $864.2 \pm 16.4$ \\
\hline
$\sigma_5$ & $2199 \pm 95$ & $3197 \pm 309$ \\
\hline
$X_0$ & $-1.51 \pm 0.22$ & $8.6 \pm 0.8$ \\ 
\hline
\hline
$\alpha$ & $1.043 \pm 0.004$ & $1.056 \pm 0.012$ \\ 
\hline \hline
\end{tabular}
\end{table}

Fig.~\ref{fig-datafit} and Table~\ref{tab-datafit} show the results 
of the lifetime difference fit to the data. The CL of the fit is 94\%. 
The fitted $D^0 \ra K^- \pi^+$ lifetime is 
$412.6 \pm 1.1$~fs, and the fitted lifetime 
difference between $D^0 \ra K^- \pi^+$ and $D^0 \ra K^+ K^-$
is $4.76 \pm 2.85$~fs. Together these values give
$y_{CP} = 1.15 \pm 0.69$\%. 

The parameters obtained by the fit for the resolution function 
(see Table~\ref{tab-datafit}) differ from those obtained from the 
MC study by more than the quoted errors. This is nominally due to 
vertex detector misalignment, IP mismeasurement, material unaccounted 
for in the MC simulation, and momentum dependence. These effects are 
included when evaluating the overall systematic error on $y^{}_{CP}$
(Table~\ref{tab-ycp-syst}).

\begin{table}[htb]
\caption{Systematic errors of the $y_{CP}$ measurement.}
\label{tab-ycp-syst}
\begin{tabular}
{@{\hspace{1.0cm}}l@{\hspace{2.0cm}} @{\hspace{1.0cm}}c@{\hspace{1.0cm}}} 
\hline \hline
Source & Value (\%) \\
\hline
$X_0(K\pi) = X_0(KK)$ assumption & 0.15 \\
\hline
$\alpha$ difference on data and MC & 0.20 \\
\hline
Bin size & 0.10 \\
\hline 
$D^0$ mass window & 0.20 \\
\hline
Background & 0.10 \\
\hline
Selection requirements & 0.15 \\
\hline \hline
Total & 0.38 \\
\hline \hline
\end{tabular}
\end{table}

\section{CPV asymmetry fit}

The CP-violating parameter \Agamma\ in $D^0 \ra K^+ K^-$
decays is obtained from fits to the data using  
the charge of the slow pion from $D^{*\pm}\ra D^0\pi^+$ or $\overline{D}{}^{\,0}\pi^-$ 
to distinguish $D^0$ decays from $\overline{D}{}^{\,0}$ decays. The $D^0$ and 
$\overline{D}{}^{\,0}$ subsamples are fit separately using the resolution function
parameters obtained from the full statistics fit. Background is treated
as described in the previous section.
The same procedure is performed for $D^0 \ra K^- \pi^+$ decays and
also for MC events in order to estimate the systematic uncertainty
of the fit. All results are listed in Table~\ref{tab-cpv}. The 
systematic uncertainty of the measurement is taken to be $\pm 0.30\%$.

\begin{table}[htb]
\caption{Results of \Agamma\ fit on data and MC.}
\label{tab-cpv}
\begin{tabular}
{@{\hspace{0.3cm}}c@{\hspace{0.3cm}} @{\hspace{0.3cm}}c@{\hspace{0.3cm}}
@{\hspace{0.3cm}}c@{\hspace{0.3cm}} @{\hspace{0.3cm}}c@{\hspace{0.3cm}}} 
\hline \hline
$D^0$ decay mode & data/MC & $\tau^{}_{(q=+1)} - \tau^{}_{(q=-1)}$
(fs) & \Agamma\ (\%) \\
\hline
$K\pi$ & MC & $1.2 \pm 1.1$ & $0.15 \pm 0.13$ \\
\hline
$K K$ & MC & $2.6 \pm 3.1$ & $0.31 \pm 0.38$ \\
\hline
$K\pi$ & data & $-0.76 \pm 1.55$ & $-0.09 \pm 0.19$ \\
\hline
$K K$ & data & $-1.67 \pm 5.18$ & $-0.20 \pm 0.63$ \\
\hline \hline
\end{tabular}
\end{table}

\section{Summary}

In conclusion, we report a preliminary measurement of the \dd\ 
mixing parameter $y_{CP}$ and the CP-violating parameter \Agamma\ 
obtained by measuring lifetime differences in $D^0\ra K^-\pi^+$ and
$D^0\ra K^+ K^-$ decays. The flavor of the $D^0$ or $\overline{D}{}^{\,0}$ 
is identified via $D^{*+}\rightarrow D^0\pi^+$ decays.
The results obtained are
$y_{CP} = (1.15 \pm 0.69 \pm 0.38)\%$ and 
$A^{}_\Gamma = (-0.20 \pm 0.63 \pm 0.30)\%$.

\section{Acknowledgement}

We wish to thank the KEKB accelerator group for the excellent
operation of the KEKB accelerator.
We acknowledge support from the Ministry of Education,
Culture, Sports, Science, and Technology of Japan
and the Japan Society for the Promotion of Science;
the Australian Research Council
and the Australian Department of Education, Science and Training;
the National Science Foundation of China under contract No.~10175071;
the Department of Science and Technology of India;
the BK21 program of the Ministry of Education of Korea
and the CHEP SRC program of the Korea Science and Engineering Foundation;
the Polish State Committee for Scientific Research
under contract No.~2P03B 01324;
the Ministry of Science and Technology of the Russian Federation;
the Ministry of Education, Science and Sport of the Republic of Slovenia;
the National Science Council and the Ministry of Education of Taiwan;
and the U.S.\ Department of Energy.

\end{document}